\documentclass[conference, a4paper]{IEEEtran}

\usepackage{siunitx}
\usepackage{pbox}
\usepackage{makecell}
\usepackage{float}
 \usepackage{epsfig}
\usepackage[T1]{fontenc}
\usepackage[utf8]{inputenc}
\usepackage{graphicx}
\usepackage[square, numbers, comma, sort & compress]{natbib}
\usepackage{amsthm}      
\usepackage{amsmath}
\usepackage{url} 
\usepackage{amssymb}
\usepackage{caption}
\usepackage{subcaption}
\usepackage{acronym} 
\usepackage{mathrsfs}
\usepackage{csquotes}
\usepackage{booktabs}
\usepackage{url}
\usepackage{xcolor}
\usepackage[ruled,vlined,noend]{algorithm2e}
\usepackage[noend]{algpseudocode}
\usepackage{multirow}
\usepackage[shortcuts,acronym]{glossaries}
\usepackage{comment}
\usepackage{mathtools}
\usepackage{afterpage} 
\usepackage{pgf}
\usepackage{color}
\usepackage[normalem]{ulem}
\usepackage{balance}
\usepackage{svg}
\usepackage[utf8]{inputenc}
\usepackage{pgfplots}
\DeclareUnicodeCharacter{2212}{−}
\usepgfplotslibrary{groupplots,dateplot}
\usetikzlibrary{patterns,shapes.arrows}
\pgfplotsset{compat=newest}

\newacronym{g2g}{G2G}{Glass-to-Glass}
\newacronym{m2m}{M2M}{Motion-to-Motion}
\newacronym{rpi}{RPI}{Raspberry Pi}
\newacronym{ntp}{NTP}{Network Time Protocol}
\newacronym{pps}{PPS}{Pulse Per Second}
\newacronym{gpio}{GPIO}{General-Purpose Input/Output}
\newacronym{cavs}{CAVs}{Connected and Autonomous Vehicles}
\newacronym{iqr}{IQR}{Interquartile Range}

\begin{document}
\title{Motion-to-Motion Latency Measurement Framework for Connected and Autonomous Vehicle Teleoperation}
\author{\IEEEauthorblockN{Fran\c{c}ois PROVOST, Faisal HAWLADER, Mehdi TESTOURI and Raphaël FRANK}
\IEEEauthorblockA{Interdisciplinary Centre
for Security, Reliability and Trust (SnT)\\
University of Luxembourg, 29 Avenue J.F Kennedy,
L-1855 Luxembourg\\
firstname.lastname@uni.lu}
}
\maketitle
\begin{abstract}
Latency is a key performance factor for the teleoperation of Connected and Autonomous Vehicles (CAVs). 
It affects how quickly an operator can perceive changes in the driving environment and apply corrective actions. 
Most of the existing work focuses on Glass-to-Glass (G2G) latency, which captures delays only in the video pipeline. 
However, there is no standard method for measuring the Motion-to-Motion (M2M) latency, defined as the delay between the physical steering movement of the remote operator and the corresponding steering motion in the vehicle. 
This paper presents an M2M latency measurement framework that uses Hall-effect sensors and two synchronized Raspberry Pi (RPI)~5 devices. The system records interrupt-based timestamps on both sides to estimate M2M latency, enabling M2M latency estimation independently of the underlying teleoperation architecture. Precision tests show a 10–15~ms accuracy, while field results indicate actuator-dominated M2M latency with medians above 750 ms.
\end{abstract}
\vspace{3px}
%
\vspace{-4px}
\section{Introduction}
\label{sec:introduction}
Teleoperation of \ac{cavs} enables a human operator to control a vehicle from a remote location and provides a critical fallback mechanism when autonomous systems reach their operational limits. 
In such scenarios, latency is one of the primary factors determining how effectively an operator can perceive the driving environment and apply corrective control actions. 
Prior studies consistently show that human performance degrades sharply as latency increases. 
For instance, \cite{chen_human_2007} recommends keeping latency below 170~ms, while \cite{kamtam_network_2024} reports that delays exceeding 300~ms can make teleoperated driving challenging. 
Beyond 400~ms \cite{neumeier_measuring_2019-1}, real-time teleoperation performance becomes difficult to maintain.

Accurate characterisation of latency is therefore essential for evaluating and improving teleoperated driving systems. 
Most existing work focuses on \ac{g2g} latency, which measures the delay from camera capture in the vehicle to the corresponding display output at the operator side. 
This delay is typically assessed using software time-stamping or light-sensor-based methods~\cite{yang_5G-nr_2022, ando_basic_2015, kroep_breaking_2025, den_ouden_design_2022-1, olabisi_evaluating_2024, neumeier_measuring_2019-1}. 
Importantly, \ac{g2g} latency captures only the perceptual portion of the end-to-end delay.
An equally important but largely unexplored component is the \ac{m2m} latency, which refers to the delay between the operator’s physical steering movement and the resulting steering motion in the vehicle, taking into account actuation.

To the best of our knowledge, no existing method in the literature provides a reliable or reproducible way to measure \ac{m2m} latency. 
A related concept, Motion-to-Photon latency, is commonly measured in virtual-reality systems~\cite{kroep_breaking_2025, seo_photosensor-based_2017}, but it captures display response time rather than vehicle actuation. 
Measuring \ac{m2m} latency is substantially more challenging because it requires detecting two physically separated mechanical events with high temporal precision while maintaining precise synchronization between the recording devices.

To address this gap, this paper introduces a novel framework for dynamically measuring \ac{m2m} latency in teleoperation of \ac{cavs}\footnote{Source code available at: \url{https://github.com/sntubix/m2m_latency.git}}.
The system employs Hall-effect sensors mounted on the steering wheels of both the remote operator station and the vehicle. 
Each sensor triggers a hardware interrupt that is timestamped on two \ac{rpi}~5 devices synchronized using Chrony\footnote{Chrony: \url{https://chrony-project.org/}}. 
The proposed framework operates independently of the teleoperation software stack and does not interfere with the control systems, making it suitable for real-world experimentation.
\begin{figure*}[!t]
    \centering
    \includegraphics[width=0.95\linewidth]{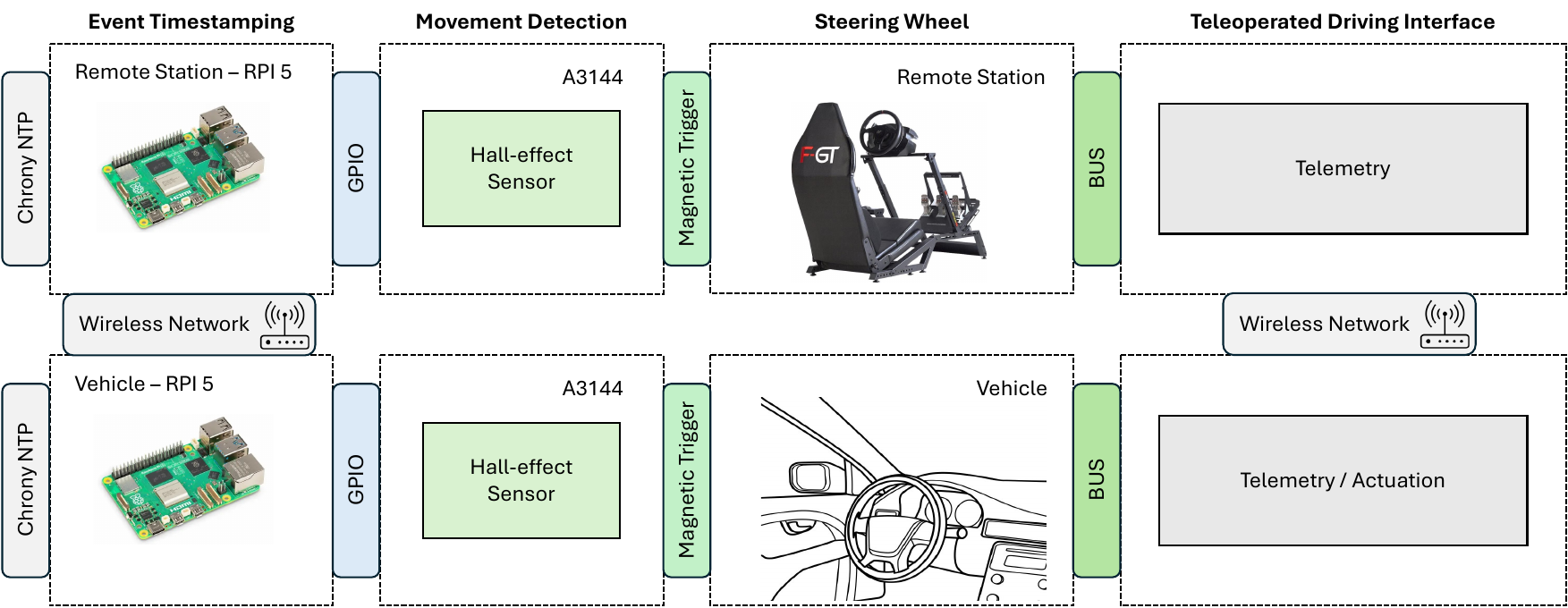}
    \caption{Overview of the \ac{m2m} latency measurement framework. Hall-effect sensors detect steering wheel motion on both sides. Each detection triggers a hardware interrupt that is timestamped on two Raspberry Pi~5 devices synchronized using Chrony.}
    \label{fig:m2m_latency_framework}
\end{figure*}

\section{Architecture}
\label{Sec:Architecture}
The overall system architecture, shown in Fig.~\ref{fig:m2m_latency_framework}, is designed to measure \ac{m2m} latency using two \ac{rpi}~5 devices with clocks synchronized via Chrony.
Each device records a hardware interrupt generated by a Hall-effect sensor, detecting magnets mounted on the steering wheel.
The first interrupt \((Event_1)\) is triggered at the remote station when the operator moves the steering wheel.
The second interrupt \((Event_2)\) is triggered on the vehicle side when the steering wheel moves in response to the actuation system.
The \ac{m2m} latency is computed as:

\begin{equation}
    M2M = Event_2 - Event_1 
    \label{eq:latencyForm}
\end{equation}

When the vehicle is stationary or moving slowly, tire-ground friction can introduce an additional delay in the steering response. 
The total measured delay is expressed as:
\begin{equation}
L_{total} = L_{gen} + L_{network} + L_{exec} + L_{follow} + E_{total}
\label{eq:latencyDistrib}
\end{equation}
where \(L_{gen}\) is the command generation delay on the operator side before the control command is transmitted, 
\(L_{network}\) is the communication latency across the network, 
\(L_{exec}\) is the command execution delay on the vehicle side before actuation begins, 
\(L_{follow}\) is the mechanical delay between actuator input and the resulting steering wheel motion, 
and \(E_{total}\) represents the combined measurement error of the system.
The error term is further decomposed as:
\begin{equation}
E_{total}=E_{sync}+E_{circuit}+E_{kernel}+E_{calib}
\label{eq:ErrorForm}
\end{equation}

where \(E_{\text{sync}}\) is the clock synchronization error between the two devices, 
\(E_{\text{circuit}}\) is the difference in signal-propagation delay from the Hall-effect sensor to the \ac{gpio} pin in both \ac{rpi} 5, 
\(E_{\text{kernel}}\) is the difference in kernel interrupt handling latency in both \ac{rpi} 5, 
and \(E_{\text{calib}}\) is the error introduced by sensor alignment and placement. 
The term \(E_{\text{circuit}} + E_{\text{kernel}}\) captures the difference between the physical occurrence of an event and the timestamp recorded on each device. 
If these delays are identical on both \ac{rpi}~5 units, their contribution does not affect the measured latency.
\subsection{Hardware}
The system uses two \ac{rpi}~5 boards equipped with 8 GB of RAM. 
Both boards run Ubuntu Server~24.04 (kernel 6.8), installed using the official Pi Imager. 
Steering wheel motion is detected using an A3144 Hall-effect sensor paired with round Neodymium disk magnets. 
A 1~k$\Omega$ pull-up resistor is connected between the sensor output and the supply voltage so that the signal remains high in the absence of a magnetic field and drops low when a magnet passes in front of the sensor. 
The sensor is powered from the 3.3~V rail of the \ac{rpi}~5 to ensure compatibility with the \ac{gpio} interrupt input.
\subsection{Software}
\label{SubSec:SoftAndTps}
The interrupt events on the \ac{gpio} pins are handled inside the Linux kernel through a custom C kernel module that is manually loaded at runtime. 
Implementation details of the kernel module can be found in our open-source repository (see footnote in Section~\ref{sec:introduction}).
The module records timestamps using the POSIX real-time clock and writes them to the kernel ring buffer. 
To reduce scheduling variability, the interrupt handler execution affinity is set to core~2 of the \ac{rpi}~5. 
Clock synchronization is performed using Chrony, which is pinned to core~0, and evaluated under two configurations:

\begin{enumerate}
    \item \textbf{Co-referenced synchronization:} One board is synchronized to the default Chrony \ac{ntp} servers and then acts as the reference clock for the second board. This configuration requires both boards to be on the same network and able to communicate.
    \item \textbf{Autonomous synchronization:} Each board is independently synchronized to the default Chrony \ac{ntp} servers. This configuration does not require a shared network and reflects typical teleoperation conditions.
\end{enumerate}
\section{Preliminary Evaluation}
This section presents the preliminary results obtained to assess the accuracy of the measurement of the proposed \ac{m2m} latency evaluation framework.
First, we describe the procedure used to quantify the intrinsic precision of the system, specifically the accuracy of clock synchronization, the kernel-level scheduling delay associated with interrupt handling, and the sensor calibration error.
These measurements define the minimum achievable timing error of the setup.
Second, we present field experiments conducted under static and dynamic teleoperation scenarios to evaluate the performance of the framework under real-world operating conditions.
\subsection{Precision Test}
\label{SubSec:SyncTests}
\subsubsection{\textbf{Experiment Setup}}
The purpose of this experiment is to evaluate the contributions of \(E_{circuit}\), \(E_{sync}\), \(E_{kernel}\), and \(E_{calib}\) to the total measurement error defined in Eq.~\ref{eq:ErrorForm}.

To assess \(E_{sync}\), a GPIO pin was toggled every 500~ms using a custom C++ application. 
The program was assigned maximum priority and pinned to core~3 of the \ac{rpi}~5 to minimize software-induced latency. 
The output of this GPIO pin was wired directly to the interrupt input on both \ac{rpi}~5 devices, enabling direct comparison of the timestamps recorded by each.
To obtain statistically significant measurements, the experiment was run continuously for one hour.

In parallel, \(E_{kernel}\) was evaluated using the Cycletest tool. Cycletest was set to monitor core 2, as we set the IRQ to have affinity for this core.
This setup ensures that the measured scheduling latency accurately reflects the delay introduced by kernel-level interrupt handling. 
Both the co-referenced and autonomous synchronization configurations described in Section~\ref{SubSec:SoftAndTps} were tested.
Given the short wiring distance and the high propagation speed of electrical signals (greater than \(2 \times 10^8\) m/s), the propagation delay through the circuit is negligible. 
As a result, the only meaningful contributor to \(E_{circuit}\) is the reaction time of the Hall-effect sensor, which is approximately 2~µs. 
Because this value is negligible compared to the other error sources, \(E_{circuit}\) is not considered further.

The calibration error \(E_{calib}\) arises from the physical placement of the Hall-effect sensor relative to the magnet on the steering wheel. 
A misalignment of about 1° is possible when attaching the sensor assembly. 
Assuming a typical steering rate of 100°/s during maneuvers, this angular offset corresponds to an uncertainty of roughly 10~ms in the detected event timing.
\begin{table}[h]
\centering
\resizebox{0.95\linewidth}{!}{
\begin{tabular}{lcccccccc}
\toprule
\multirow{2}{*}{\textbf{Mode}} &
\multicolumn{4}{c}{\textbf{Synchronization Offset (ms)}} &
\multicolumn{4}{c}{\textbf{Scheduling Latency (ms)}} \\
\cmidrule(lr){2-5} \cmidrule(lr){6-9}
& Min & Max & Mean & Std & \textbf{RPI} & Min & Max & Mean \\
\midrule
\multirow{2}{*}{Co-Ref} 
& 0.000256 & 4.446 & 0.322 & 0.468 & 1 & 0.002 & 0.062 & 0.005 \\
& & & & & 2 & 0.003 & 0.052 & 0.005 \\
\midrule
\multirow{2}{*}{Auto} 
& 0.000258 & 1.052 & 0.330 & 0.219 & 1 & 0.002 & 0.118 & 0.005 \\
& & & & & 2 & 0.002 & 0.106 & 0.005 \\
\bottomrule
\end{tabular}
}
\caption{\small Synchronization offset and kernel scheduling latency measured under co-referenced (Co-Ref) and autonomous (Auto) clock synchronization configurations.}
\label{tab:Synchronization_and_Scheduling}
\end{table}
\subsubsection{\textbf{Precision Test Results}}
The results of the precision tests are summarized in Table~\ref{tab:Synchronization_and_Scheduling}. 
The table reports two categories of measurements: the synchronization offset between the two devices and the kernel scheduling latency associated with interrupt handling.

\textbf{Synchronization Offset:}
For both synchronization modes, the mean offset remains close to 0.3~ms, indicating that the two \ac{rpi}~5 boards maintain a closely aligned notion of time. 
The minimum offset values (\(\sim\)0.00026~ms) show that the clocks can occasionally achieve near-perfect alignment, while the maximum values capture brief deviations caused by network jitter or the timing of Chrony correction updates. 
The wider maximum range observed in the co-referenced configuration (up to 4.4~ms) is attributed to occasional fluctuations introduced when one \ac{rpi} acts as the reference for the other. 
Despite this, the low mean and standard deviation values confirm that both synchronization modes are reliable for our latency measurement purpose.

\textbf{Kernel Scheduling Latency:}
The second part of Table~\ref{tab:Synchronization_and_Scheduling} reports the interrupt scheduling latency measured by Cycletest. 
Both devices exhibit nearly identical behavior, with mean values around 5~µs. 
The maximum observed scheduling latencies are slightly higher in the autonomous configuration, but still remain below 0.12~ms. 
Since \ac{m2m} latency is computed as the difference between two timestamps, and both boards introduce similar interrupt handling delays, these kernel latencies are mostly negligible. 
Even in the worst case, where one board experiences its maximum scheduling delay while the other experiences its minimum, the resulting asymmetry is on the order of 100~µs, which is also negligible.

\textbf{Key Findings:}
The combination of a synchronization accuracy of around 0.3~ms and µs-level kernel scheduling jitter confirms that the intrinsic timing precision of the proposed framework is high. 
When the calibration error (\(\sim10\)~ms) is included, the overall measurement accuracy of the system is estimated to be approximately 10--15~ms, which is sufficient for reliable \ac{m2m} latency estimation in teleoperation scenarios.

\subsection{Field Test}
\subsubsection{\textbf{Experiment Setup}}
\begin{figure}[t]
    \centering
    \subfloat[Static Scenarios]{\includegraphics[width=0.44\linewidth, trim=2 5 5 15, clip]{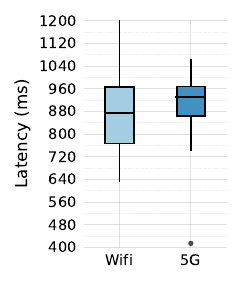}
    \label{fig:lat_static}}
    \subfloat[Dynamic Scenarios]{\includegraphics[width=0.44\linewidth, trim=2 5 5 15, clip]{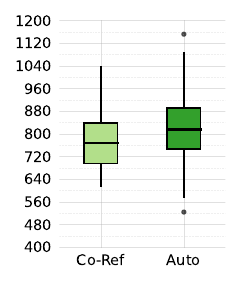}
    \label{fig:lat_dynamic}}
        \caption{\small Latency measurements for (a) static scenarios using WiFi and 5G under stationary conditions, and (b) dynamic scenarios comparing co-referenced (Co-Ref) and autonomous (Auto) synchronization while driving at approximately 10~km/h using a 5G connection.}
\label{fig:latency_results}
\end{figure}

The remote station and vehicle configuration used in these tests follows the setup described in \cite{testouri2025robocar, testouri_5g-enabled_2025}. 
The field evaluation was conducted across four scenarios, grouped into two categories static and dynamic, as illustrated in Fig.~\ref{fig:latency_results}. 
All tests were carried out in a parking lot located in Kirchberg in Luxembourg\footnote{Test route: \url{http://g-o.lu/3/WYZn}}. 
The static tests were performed using the co-referenced synchronization and evaluated two network technologies: WiFi, representing an ideal low-latency baseline, and a commercial 5G network from POST Luxembourg, representing a more realistic deployment scenario. 
In both static cases, the vehicle remained stationary while only the steering wheel was actuated.

The dynamic tests compared the co-referenced and autonomous synchronization. 
Both tests were conducted using the commercial 5G network while the vehicle was driven at a speed of $\sim$10~km/h over a short loop of about 100~m. 
\subsubsection{\textbf{Field Test Results}}
Fig.~\ref{fig:latency_results} summarises the latency distributions in all four field-test scenarios. 
Across all cases, the observed median latency exceeds 750~ms. 
This behaviour is consistent with the findings reported in~\cite{testouri_5g-enabled_2025}, where the dominant contribution to the end-to-end delay originates from the vehicle’s steering actuator and its PID controller. 

\textbf{Static Scenarios (WiFi vs. 5G):}
In the static configuration, WiFi produces a median latency of 874.5~ms with an \ac{iqr} of 198.0~ms. 
Static 5G results in a slightly higher median latency of 930.6~ms but a substantially narrower \ac{iqr} of 105.0~ms (a reduction of $\sim$47\%). 

The standard deviation also decreases from 126.8~ms (WiFi) to 95.8~ms (5G).  
These results indicate that although WiFi achieves a lower median latency value, 5G provides more stable and predictable latency with fewer high delay outliers. 
Such reduced variability is beneficial for closed-loop teleoperation, where timing consistency often outweighs small differences in median delay.

\textbf{Dynamic Scenarios (Co-Ref vs.\ Auto Synchronization):}
When the vehicle is in motion, Co-Referenced Synchronization yields a median latency of 767.8~ms, an \ac{iqr} of 141.7~ms, and only 1.4\% of samples exceeding 1~s. 
Under the same conditions, Autonomous Synchronization results in a higher median latency of 815.2~ms, a slightly larger \ac{iqr} of 145.9~ms, and a higher proportion of delays above 1~s (5.4\%).  
Overall, Co-Ref Sync demonstrates both a lower median latency and lower variability than Auto Sync. 
This indicates that the co-referenced mechanism remains more robust under low-speed mobility, likely due to fewer synchronization adjustments and reduced sensitivity to short-term clock drift.

\textbf{Cross-Scenario Comparison:}
Comparing static and dynamic results under 5G reveals that dynamic Co-Ref Synchronization reduces the median latency by $\sim$163~ms relative to static 5G. 
This reduction is attributed not to communication effects but to vehicle motion decreasing mechanical resistance in the steering system, thereby reducing actuator-induced delay.
Although actuator dynamics dominate the absolute latency values, the results clearly show that:
(1) static 5G produces significantly more stable latency than static WiFi, and  
(2) dynamic Co-Ref Sync outperforms Auto Sync in both median latency and variability.  
These patterns demonstrate that networking and synchronization choices remain observable and impactful despite the actuator imposed latency floor.

\section{Conclusion and Future Work}
This paper presented a framework for measuring \ac{m2m} latency for teleoperation of \ac{cavs} using Hall-effect sensors. 
The precision tests show that the dominant source of measurement error is the sensor calibration offset ($\sim$10~ms), while the clock-synchronisation error (mean $\sim$0.3~ms) and kernel scheduling jitter (mean $\sim$5~\textmu s) contribute negligibly. 

As a result, the overall measurement precision of the framework is in the range of 10--15~ms.
Field experiments showed that \ac{m2m} latency is dominated by the steering actuator, with median values above 750~ms across all test scenarios. In the static tests, WiFi reached 874.5~ms, while static 5G increased the median to 930.6~ms but reduced variability by 47\%. 
In the dynamic tests, Co-Referenced Synchronization achieved lower latency (767.8~ms) and fewer high-delay outliers than Autonomous Synchronization (815.2~ms). 
These results confirm that, despite the actuator-imposed delay floor, the choice of synchronization mode still influences overall latency stability.

Considering both, the measured M2M delay and typical G2G latency values reported in prior work~\cite{testouri_5g-enabled_2025}, the overall operator–vehicle responsiveness can approach around one second, which may challenge real-time teleoperation requirements.
This highlights the importance of evaluating \ac{m2m} latency independently from video latency when assessing teleoperation performance.
Future work will address two main limitations of the current framework: the $\sim$10~ms calibration uncertainty of the magnetic sensor, which may be reduced using high-rate inertial sensing, and the lack of detailed latency decomposition. 
The framework will also be extended with \ac{g2g} measurement to characterise the full operator-to-vehicle-to-operator latency loop for teleoperation.

\section*{Acknowledgments}
This work is supported by the Luxembourg National Research Fund under the 5G BRIDGES/2023-Phase 2/IS/19101381/5GDrive project.
\bibliographystyle{IEEEtran}

\end{document}